\documentclass[preprint]{elsarticle}

\usepackage{subfigure}
\usepackage{psfrag}
\usepackage{bm}
\usepackage{graphicx}
\usepackage{amsmath}
\usepackage{amsfonts}
\usepackage[version=4]{mhchem}
\usepackage{siunitx}
\usepackage{longtable,tabularx}
\usepackage{xcolor}
\usepackage{overcite}
\usepackage[numbers]{natbib}

\begin{document}
\begin{frontmatter}


\title{Adaptive Guidance and Integrated Navigation with Reinforcement Meta-Learning\tnoteref{mytitlenote}}
\tnotetext[mytitlenote]{Article partially presented as: Gaudet, B. and Linares, R. “Adaptive Guidance with Reinforcement Meta-Learning,” AAS 19-293, 29th AAS/AIAA Space Flight Mechanics Meeting, Ka'anapali, HI, January 13-17, 2019.}

\author[add1]{Brian Gaudet}
\ead{briangaudet@mac.com}
\author[add3]{Richard Linares\corref{cor1}}
\ead{linaresr@mit.edu}
\author[add1,add2]{Roberto Furfaro}
\ead{robertof@email.arizona.edu}

\address[add1]{Department of Systems and Industrial Engineering, University of Arizona, 1127 E. James E. Roger Way, Tucson, Arizona, 85721,USA}
\address[add2]{Department of Aerospace and Mechanical Engineering, University of Arizona}
\address[add3]{Department of Aeronautics and Astronautics, Massachusetts Institute of Technology, Cambridge, MA 02139, USA}
\cortext[cor1]{Corresponding author}

%
%
%
%
%
%
%
%
%

\begin{abstract}
This paper proposes a novel adaptive guidance system developed using reinforcement meta-learning with a recurrent policy and value function approximator. The use of recurrent network layers allows the deployed policy to adapt real time to environmental forces acting on the agent.  We compare the performance of the DR/DV guidance law, an RL agent with a non-recurrent policy, and an RL agent with a recurrent policy in four challenging environments with unknown but highly variable dynamics. These tasks include a safe Mars landing with random engine failure and a landing on an asteroid with unknown environmental dynamics. We also demonstrate the ability of a RL meta-learning optimized policy to implement a guidance law using observations consisting of only Doppler radar altimeter readings in a Mars landing environment, and LIDAR altimeter readings in an asteroid landing environment, thus integrating guidance and navigation.
\end{abstract}


\end{frontmatter}

%

\parindent=0.5 cm

\section{Introduction}

Many space missions take place in environments with complex and time-varying dynamics that may be  incompletely modeled during the mission design phase. For example, during an orbital refueling mission, the inertia tensor of each of the two spacecraft will change significantly as fuel is transferred from one spacecraft to the other, which can make the combined system difficult to control [\citenum{guang2018attitude}]. A significant fraction of an exoatmospheric kill vehicle's (EKV) wet mass consists of fuel, and as this is depleted with divert thrusts, the center of mass changes,  which impacts performance. Future missions to asteroids might be undertaken before the asteroid's gravitational field, rotational velocity, and local solar radiation pressure are accurately modeled. Also consider that it is difficult to accurately model the aerodynamics of hypersonic re-entry. Moreover, there is the  problem of navigation system errors biasing the state estimate given to the guidance system. Finally, there is the possibility of actuator failure, which significantly modifies the dynamic system of a spacecraft and its environment. These examples show a clear need for a guidance system that can adapt in real time to time-varying system dynamics that are likely to be imperfectly modeled prior to the mission. The integration of guidance and navigation bring further benefits, including the ability to adapt to sensor distortion, and co-optimization of the two systems.

Recent work in the area of adaptive guidance algorithms include [\citenum{prabhakar2018trajectory}], which demonstrates an adaptive control law for a UAV tracking a reference trajectory, where the adaptive controller adapts to external disturbances. One limitation is the linear dynamics model, which may not be accurate, as well as the fact that the frequency of the disturbance must be known. In [\citenum{huang2019mars}], the authors develop a fault identification system for Mars entry phase control using a pre-trained neural network, with a fault controller implemented as a second Gaussian neural network,  Importantly, the second network requires on-line parameter update during the entry phase, which may not be possible to implement in real time on a flight computer. Moreover, the adaptation is limited to known actuator faults as identified by the 1st network. In [\citenum{han2015adaptive}],the authors develop an adaptive controller for spacecraft attitude control using reaction wheels. This approach is also limited to actuator faults, and the architecture does not adapt to either state estimation bias or environmental dynamics.

Several works have demonstrated improved performance with uncertain and complex dynamics using the reinforcement learning (RL) framework and training with randomized system parameters. In [\citenum{yu2017preparing}], the authors use a recurrent neural network to explicitly learn model parameters through real time interaction with an environment; these parameters are then used to augment the observation for a standard reinforcement learning algorithm.   In  [\citenum{peng2017sim}], the authors use a recurrent policy and value function in a modified deep deterministic policy gradient algorithm to learn a policy for a robotic manipulator arm that uses real camera images as observations. In both cases, the agents train over a wide range of randomized system parameters.  In the deployed policy, the recurrent network's internal state quickly adapts to the actual system dynamics, providing good  performance for the agent's respective tasks.

In this work we use the principles of reinforcement meta-learning [\citenum{wang2016learning}] (meta-RL) to formulate two adaptive guidance laws, one suitable for controlling a lander in a Mars powered descent phase, the other suitable for landing on small bodies such as an asteroid. In the meta-RL framework, and agent "learns to learn" through exposure to a multitude of environments.  The optimized policy can then quickly adapt to novel environments.  The guidance laws take the form of  a global policy over the region of state space defined by the deployment region and potential landing sites. This global policy maps the navigation system's estimate of the lander's state to a commanded thrust vector in the target centered reference frame.  Optimizing the policy involves simulated interaction between an agent instantiating the policy and the environment over many episodes with randomly generated initial conditions that cover the theater of operations. Importantly, environmental parameters such as state estimation error, the lander's wet mass, and environmental forces are varied between episodes. The optimized policy will adapt in real time to these  parameters.  We will demonstrate through a series of experiments that the meta-RL adaptive guidance law outperforms  a traditional energy-optimal closed-loop guidance algorithm independently developed by  by Battin [\citenum{battin1999introduction}, page 558] and D'Souza [\citenum{d1997optimal}], referred to in the sequel as a "DR/DV policy".  Moreover, we demonstrate that in a subset of experiments, the meta-RL adaptive guidance law (referred to in the sequel as the "meta-RL policy") outperforms a non-adaptive guidance law optimized using standard reinforcement learning (hereafter referred to as an "RL policy"). In the Mars landing experiments, the RL policy is identical to the 3-DOF Mars landing policy developed in [\citenum{gaudet2018deep}]. Finally, in two experiments, we integrate guidance and navigation by optimizing a guidance law that takes observations consisting of sensor outputs. The first experiment is in a Mars powered descent phase environment, where the observation consists of simulated Doppler radar altimeter readings. The second experiment is in an asteroid landing environment using simulated LIDAR altimeter readings. 

We compare the performance of the DR/DV policy,  RL policy, and a meta-RL  policy over a range of tasks with unknown but highly variable dynamics. We use the DR/DV policy  as a performance baseline, and to improve its performance, we give it access to the ground truth lander mass at the start of an episode. In contrast, the RL and meta-RL policies only have access to observations that are a function of the lander's position and velocity, or in some cases, sensor readings. Since sensor output feedback is still an open problem in the optimal control framework, for the tasks using sensor output as observations, we did not compare performance to that obtainable using DR/DV.  These tasks include:

\begin{enumerate}
    \item \underline{Engine Failure (3-DOF Mars Landing):}  At the start of each episode, with probability $p$ a the lander's thrust capability in a given direction is reduced.
    \item \underline{Large Mass Variation (3-DOF Mars Landing):} We use a small engine specific impulse and assume wet/dry masses of 2000kg/200kg respectively, which results in a large variation in lander mass during the landing. This creates a difficult control problem, as the agent does not have access to the ground truth mass.  
    \item \underline{Landing with Integrated Guidance and Navigation:}  The agent learns a guidance law with observation consisting of the readings from the four Doppler radar altimeters. This example uses a 3-DOF Mars Landing model.
    \item \underline{State Estimate Bias: (3-DOF Mars Landing):} The agent has access to a corrupted version of the ground truth state. 
    \item \underline{Unknown Dynamics (3-DOF Asteroid Landing):}  In each episode, the acceleration due to gravity, solar radiation pressure, and rotation are randomly chosen over a wide range, limited only by the lander's thrust capability.
    \item \underline{Landing with Integrated Guidance and Navigation:} The agent learns a guidance law with observation consisting of LIDAR altimeter readings. This example uses a 3-DOF Asteroid Landing model. 
    
\end{enumerate}

\section{Meta-RL Problem Formulation}

In the RL framework, an agent learns through repeated interaction with an environment how to complete a task. A Markov Decision Process (MDP) is an abstraction of this environment, which in a continuous state and action space, can be represented by a state space $\mathcal{S}$, an action space $\mathcal{A}$, a state transition distribution $\mathcal{P}(\mathbf{x}_{t+1}|\mathbf{x}_t,\mathbf{u}_t)$, and a reward function $r=\mathcal{R}(\mathbf{x}_t,\mathbf{u}_t))$, where $x \in \mathcal{S}$ and $u \in \mathcal{A}$, and $r$ is a scalar reward signal. We can also define a partially observable MDP (POMDP), where the state $\mathbf{x}$ becomes a hidden state, generating an observation $\mathbf{o}$ using an observation function $\mathcal{O}(\mathbf{x})$ that maps states to observations. The POMDP formulation is useful when the observation is a noisy version of the ground truth state (due to navigation system inaccuracies) and when the observation consists of raw sensor outputs such as LIDAR readings or sequential camera images.  In the following, we will refer to both fully observable and partially observable environments as POMDPs, as an MDP can be considered a POMDP with an identify function mapping states to observations.

The agent operates within an  environment defined by the POMDP, generating some action $\mathbf{u}_t$ based off of the observation $\mathbf{o}_t$, and receiving reward $r_{t+1}$ and next observation $\mathbf{o}_{t+1}$. Optimization involves maximizing the sum of (potentially discounted) rewards over the trajectories induced by the interaction between the agent and environment. Constraints such as minimum and maximum thrust, glide slope, attitude compatible with sensor field of view,  maximum rotational velocity, and terrain feature avoidance (such as targeting the bottom of a deep crater) can be included in the reward function, and will be accounted for when the policy is optimized. Note that there is no guarantee on the optimality of trajectories induced by the policy, although in practice it is possible to get close to optimal performance by tuning the reward function. 

Reinforcement meta-learning differs from generic reinforcement learning in that the agent learns to adapt to novel POMPDs by learning over a wide range of POMDPs. These POMDPs can include different environmental dynamics, actuator failure scenarios, and varying amounts of navigation system state estimation bias. Learning within the RL meta-learning framework results in an agent that can quickly adapt to novel POMDPs, often with just a few steps of interaction with the environment. There are multiple approaches to implementing meta-RL.  In  [\citenum{finn2017model}], the authors design the objective function to explicitly to make the model parameters transfer well to new tasks. In [\citenum{mishra2018simple}], the authors demonstrate state of the art performance using temporal convolutions with soft attention. And in [\citenum{frans2017meta}], the authors use a hierarchy of policies to achieve meta-RL. In this proposal, we use a different approach similar to that employed in  [\citenum{wang2016learning}] using a recurrent policy and value function. Note that it is possible to train over a wide range of POMDPs using a non-meta RL algorithm, and we demonstrated in [\citenum{gaudet2018deep}] that this can provide a robust integrated guidance and control system for the 6-DOF Mars landing problem, and such an approach was applied in [\citenum{rajeswaran2016epopt}] to obtain robust results in a the openAI mujoco environment with robot parameters randomly varied across episodes. However, although such an approach typically results in a robust policy, the policy cannot adapt in real time to novel environments. 

In this work the meta-RL policy is optimized using proximal policy optimization (PPO) [\citenum{schulman2017proximal}] with both the policy and value function implementing recurrent layers in their networks.  To understand how recurrent layers result in an adaptive agent, consider that given some ground truth agent position and velocity $\mathbf{x}\in\mathbb{R}^{6}$, and action vector $\mathbf{u}\in\mathbb{R}^{3}$ output by the agent's policy, the next observation depends not only on $\mathbf{x}$ and $\mathbf{u}$, but also on the ground truth agent mass and external forces acting on the agent. Consequently, during training, the hidden state of a policy's recurrent layers evolves differently depending on the observed sequence of observations from the environment and actions output by the policy. Specifically, the trained policy's hidden state captures unobserved information such as (potentially time-varying) external forces that are useful in minimizing the cost function. In contrast, a non-recurrent policy, which does not maintain a persistent state vector, can only optimize using a set of current observations, actions, and advantages, and will tend to under-perform a recurrent policy on tasks with randomized dynamics.  After training, although the recurrent policy's network weights are frozen, the hidden state will continue to evolve in response to a sequence of observations and actions, thus making the policy adaptive. In contrast, an non-recurrent policy's behavior is fixed by the network parameters at test time.

The policy and value functions are implemented using four layer neural networks with tanh activations on each hidden layer. Layer 2 for the policy and value function is a recurrent layer implemented as a gated recurrent unit [\citenum{chung2015gated}]. The network architectures are as shown in Table \ref{tab:NN}, where $n_{\mathrm{hi}}$ is the number of units in layer $i$, $\mathrm{obs\_dim}$ is the observation dimension, and $\mathrm{act\_dim}$ is the action dimension. During training, the hidden layer is unrolled for a number of steps in the forward pass in order to perform back-propagation through time, which allows the network to learn temporal dependencies in the rollout trajectories.

\begin{table}[h]
	\fontsize{8}{8}\selectfont
    \caption{Policy and Value Function network architecture}
   \label{tab:NN}
        \centering 
   \newcolumntype{R}{>{\raggedleft\arraybackslash}p{1.8cm}}
   \begin{tabular}{l | R | c | R | c } 
      \hline 
       & \multicolumn{2}{c}{Policy Network}\vline & \multicolumn{2}{c}{Value Network}\\
       \hline
       Layer & \# units & activation & \# units & activation \\
       \hline
      hidden 1      & $10 * \mathrm{obs\_dim}$ & tanh & $10 * \mathrm{obs\_dim}$ & tanh \\
      hidden 2      & $\sqrt{n_{\mathrm{h1}} * n_{\mathrm{h3}}}$ & tanh & $\sqrt{n_{\mathrm{h1}} * n_{\mathrm{h3}}}$ & tanh\\
      hidden 3      & $10 * \mathrm{act\_dim}$ & tanh & 5 & tanh \\
      output        & $\mathrm{act\_dim}$ & linear & 1 & linear \\
      \hline
   \end{tabular}
\end{table}

We optimize the policy using proximal policy optimization (PPO) [\citenum{schulman2017proximal}].  Our implementation of the PPO algorithm uses an advantage function that is the difference between the empirical return and a state value function baseline, as shown  in Eq. \eqref{eq:ppo_adv}, where $r(\mathbf{u},\mathbf{o})$ is the reward function, and $\gamma$ a discount rate used to facilitate temporal credit assignment:

\begin{equation}
\label{eq:ppo_adv}
	A^{\pi}_{\bf w}(\mathbf{o}_{k},\mathbf{u}_{k})=\left[\sum_{\ell=k}^{T}\gamma^{\ell-k}r(\bf o_{\ell},\bf u_{\ell})\right]-V_{\bf w}^{\pi}(\mathbf{o}_{k})
\end{equation}
Here the value function $V_{\bf w}^{\pi}$ is learned using the cost function given  in Eq.  \eqref{eq:vf_ppo}.
\begin{equation}
\label{eq:vf_ppo}
L(\mathbf{w})=\sum_{i=1}^{M}\left(V_{\mathbf{w}}^{\pi}({\bf o}_k^i)-\left[\sum_{\ell=k}^{T}\gamma^{\ell-k}r({\bf u}_{\ell}^i,{\bf o}_{\ell}^i)\right]\right)^2
\end{equation}
PPO attempts to bound changes in the policy's distribution of actions conditioned on observations using a probability ratio heuristic $p_{k}({\bm \theta})$ given by,
\begin{equation}
\label{eq:clipr}
p_{k}({\bm \theta})=\frac{\pi_{{\bm \theta}}({\bf u}_{k}|{\bf o}_{k})}{\pi_{{\bm \theta}_\text{old}}({\bf u}_{k}|{\bf o}_{k})}
\end{equation}
Our objective function is then shown in Eq. \ref{eq:ppoloss}:
\begin{equation}
\label{eq:ppoloss}
L({\bm \theta})=\mathbb{E}_{p({\bm \tau})}\left[\mathrm{min}\left[p_{k}({\bm \theta}) ,
\mathrm{clip}(p_{k}({\bm \theta}) , 1-\epsilon, 1+\epsilon)\right]A^{\pi}_{\bf w}({\bf o}_{k},{\bf u}_{k})\right]
\end{equation}
This clipped objective function has been shown to aid convergence by insuring that the policy does not change drastically between updates.

The policy optimization algorithm updates the policy using a batch of trajectories (roll-outs) collected by interaction with the environment. Each trajectory is associated with a single episode, with a sample from a trajectory collected at step $k$ consisting of observation ${\bf o}_{k}$, action ${\bf u}_{k}$, and reward $r_k({\bf o}_k,{\bf u}_k)$. Finally, gradient accent is performed on ${\bm \theta}$ and gradient decent on ${\bf w}$ and update equations are given by 
\begin{align}\label{loss}
{\bf w}^+&={\bf w}^--\beta_{{\bf w}}\nabla_{{\bf w}} \left. L({\bf w})\right|_{{\bf w}={\bf w}^-}\\
{\bm \theta}^+&={\bm \theta}^-+\beta_{{\bm \theta}} \left. \nabla_{\bm \theta}J\left({\bm \theta}\right)\right|_{{\bm \theta}={\bm \theta}^-}
\end{align}
where $\beta_{{\bf w}}$ and $\beta_{{\bm \theta}}$ are the learning rates for the value function, $V_{\bf w}^{\pi}$, and policy, $\pi_{\bm \theta}\left({\bf u}_k|{\bf o}_k\right)$, respectively.
 
In our implementation, we dynamically adjust the clipping parameter $\epsilon$ to target a KL divergence between policy updates of 0.001. The policy and value function are learned concurrently, as the estimated value of a state is policy dependent. We use a Gaussian distribution with mean $\pi_{\bm \theta}({\bf o}_{k})$ and a diagonal covariance matrix for the action distribution in the policy.  Because the log probabilities are calculated using the exploration variance, the degree of exploration automatically adapts during learning such that the objective function is maximized. 

\section{Mars Landing Environment}

Each episode begins with the initial conditions shown in Table \ref{tab:IC}, where values are in the target centered reference frame.  At the start of each episode, the environmental disturbance acceleration $\mathbf{a}_{\text{env}}$ in Eq. \ref{eq:EQOM1b} is randomly set over a uniform distribution ranging from $\text{-0.2m/s}^2\text{to 0.2m/s}^2$, and the lander's initial mass $m$ in Eq. \ref{eq:EQOM1b} is set over  a random distribution +/-10\% of nominal. These are the types and amounts of parameter variation we would expect any guidance law to be able to easily compensate for.

\begin{table}[h]
	\fontsize{10}{10}\selectfont
    \caption{Mars Lander Initial Conditions for Optimization}
   \label{tab:IC}
        \centering 
   \newcolumntype{R}{>{\raggedleft\arraybackslash}p{1.8cm}}
   \begin{tabular}{l | R | R  } 
       Parameter & min  & max  \\
       \hline
      Downrange Position (m)      &  0 & 2000\\
      Crossrange Position (m)       & -1000 & 1000 \\
      Elevation Position (m)     & 2300 & 2400 \\
      Downrange Velocity (m/s)      & -70 & -10 \\
      Crossrange Velocity (m/s)       & -30  & 30  \\
      Elevation Velocity (m/s)    & -90 & -70  \\
      Lander Mass (kg)  &   1800 & 2200 \\
   \end{tabular}
\end{table}

\subsection{Dynamics Model}

We model the Mars landings in 3-DOF, where the translational motion is modeled as shown in Eqs. \ref{eq:EQOM1a} through \ref{eq:EQOM1c}:
\begin{subequations}
\begin{align}
	\Dot{\mathbf r} &= {{\mathbf v}}\label{eq:EQOM1a}\\
	\Dot{\bf v} &= \frac{\bf T}{m}  + \mathbf{a}_{\text{env}} + \mathbf{g} \label{eq:EQOM1b}\\
	\Dot{m} &= -\frac{\lVert{{\bf T}}\rVert}{I_\text{sp}g_\text{ref}} \label{eq:EQOM1c}
\end{align}
\end{subequations}
Here $\mathbf{r}$ is the lander's position in the target centered reference frame, ${{\bf T}}$ is the lander's thrust vector  $g_\text{ref}=9.8$ $\text{m}/\text{s}^{2}$,  ${\bf g}=\begin{bmatrix} 0 & 0 & -3.7114\end{bmatrix} \text{m}/\text{s}^2$,    $I_\text{sp}=225$ s, and the lander's mass is $m$.  ${\bf a}_\text{env}$ is a vector representing environmental disturbances such as wind and variations in atmospheric density.  The minimum and maximum thrust is constrained to be in the range [2000N, 15000N], and the lander's nominal wet mass is 2000kg.  The equations of motion are integrated using 4\textsuperscript{th} order Runge-Kutta integration with a 0.05s step size, and the navigation period is 0.2s.

\subsection{Reward Function}
We use an approach similar to that taken in [\citenum{gaudet2018deep}] for our reward function.  This reward function is a combination of shaping rewards [\citenum{ng2003shaping}] and a  terminal reward. The shaping rewards give the agent hints at every time step as to how well it is performing. These hints take the form of a target velocity aligned with the line of sight vector. Specifically, we define the piece wise velocity field given in Equations \eqref{eq:vtarg1a}, \eqref{eq:vtarg1b}, \eqref{eq:vtarg1c}, \eqref{eq:vtarg1d}, and \eqref{eq:vtarg1e}, where $\tau_{1}$ and $\tau_{2}$ are hyperparameters  and $v_{o}$ is set to the magnitude of the lander's velocity at the start of the powered descent phase. We see that the shaping rewards take the form of a velocity field that maps the lander's position to a target velocity. In words, we target a location 15m above the desired landing site and target a z-component of landing velocity equal to -2m/s. Below 15m, the downrange and crossrange velocity components of the target velocity field are set to zero, which encourages a vertical descent. 
\begin{subequations}
\begin{align}
    {\bf v}_{targ}&=-v_{o}\left(\frac{{\bf \hat{r}}}{\lVert{\bf \hat{r}}\rVert}\right)\left(1-\exp\left(-\frac{t_{go}}{\tau}\right)\right)\label{eq:vtarg1a}\\
    t_{go}&=\frac{\lVert{\bf \hat{r}}\rVert}{\lVert{\bf \hat{v}}\rVert}\label{eq:vtarg1b} \\
    {\bf \hat{r}}&=\begin{cases}{\bf r} - \begin{bmatrix}0 & 0 & 15\end{bmatrix}, & \text{if } r_{2} > 15 \\ \begin{bmatrix}0 & 0 & r_{2}\end{bmatrix}, &\text{otherwise}\end{cases}\label{eq:vtarg1c} \\
    {\bf \hat{v}}&=\begin{cases}{\bf v} - \begin{bmatrix}0 & 0 & -2\end{bmatrix}, & \text{if } r_{2} > 15 \\ {\bf v}-\begin{bmatrix}0 & 0 & -1\end{bmatrix}, &\text{otherwise}\end{cases}\label{eq:vtarg1d} \\
    \tau&=\begin{cases} \tau_{1}, & \text{if } r_{2} > 15 \\ \tau_{2}, &\text{otherwise}\end{cases}\label{eq:vtarg1e} 
\end{align}
\end{subequations}

Finally, we provide a terminal reward bonus when the lander reaches an altitude of zero, and the terminal position, velocity, and glideslope are within specified limits. The reward function is then given by Equation \eqref{eq:reward_func}, where the various terms are described in the following:

\begin{enumerate}
    \item $\alpha$ weights a term penalizing the error in tracking the target velocity.
    \item $\beta$ weights a term penalizing control effort.
    \item $\gamma$ is a constant positive term that encourages the agent to keep making progress along the trajectory.
    \item $\eta$ is a bonus given for a successful landing, where terminal position, velocity, and glideslope are all within specified limits.  The limits are $r_{lim}=5$ $\text{m}$, $v_{lim}=2$ $\text{m/s}$,  and $\text{gs}_{lim}=79$ deg. The minimum glideslope at touchdown insures the lander's velocity is directed predominatly downward.
\end{enumerate}
\begin{equation}
 \begin{aligned}
 \label{eq:reward_func}
r &= \alpha\|{\bf v}-{\bf v}_{targ}\|+
\beta\|{\bf T}\|+\gamma\\
&+\eta({\bf r}_{2}<0 \ \mathrm{and}\  
\|{\bf r}\|<r_{lim} \ \\
&\mathrm{and}\  \|{\bf v}\|<v_{lim} \ \mathrm{and} \ gs > gs_{lim})\
\end{aligned}
\end{equation}
This reward function allows the agent to trade off between tracking the target velocity given in Eq.~\eqref{eq:vtarg1a}, conserving fuel, and maximizing the reward bonus given for a good landing.  Note that the constraints are not hard constraints such as might be imposed in an optimal control problem solved using collocation methods. However, the consequences of violating the constraints (a large negative reward and termination of the episode) are sufficient to insure they are not violated once learning has converged. Hyperparameter settings and coefficients used in this work are given in Table~\ref{tab:HPS}, where $\|{\bf v}_{o}\|$ is the magnitude of the lander's velocity at the start of an episode.

\begin{table}[h]
	\fontsize{10}{10}\selectfont
    \caption{Hyperparameter Settings}
   \label{tab:HPS}
        \centering 
        \newcolumntype{R}{>{\raggedleft\arraybackslash}p{1.8cm}}
   \begin{tabular}{ c | c } 
      \hline
      Hyperparameter & Value\\
      $v_{o}$ (m/s) & $\|{\bf v}_{o}\|$\\
      $\tau_{1}$ (s) & 20\\
      $\tau_{2}$ (s) & 100\\
      $\alpha$  & -0.01\\
      $\beta$   &  -0.05\\
      $\gamma$ & 0.01\\
      $\eta$ & 10 \\
   \end{tabular}
\end{table}
As shown in Eq.~\eqref{eq:obs}, the observation given to the agent during learning and testing is ${\bf v}_\text{error}={\bf v}-{\bf v}_\text{targ}$, with ${\bf v}_\text{targ}$ given in Eq.~\eqref{eq:vtarg1a} , the lander's estimated altitude, and the time to go.  Note that aside from the altitude, the  lander translational coordinates do not appear in the observation. This results in a policy with good generalization in that the policy's behavior can extend to areas of the full state space that were not experienced during learning.

\begin{equation}
\label{eq:obs}
{\text{obs}}=\begin{bmatrix}{\bf v}_\text{error}  & r_{2} & t_\text{go}\end{bmatrix}
\end{equation}

\section{Asteroid Landing Environment}

The lander's initial conditions in the asteroid environment are shown in Table \ref{tab:exp1_ic}, where the values are in the target centered reference frame. Each episode starts with a random heading error  for the lander's velocity between -45 and 45 degrees defined with respect to the lander's line of sight at the start of the landing phase. The variation in initial conditions is typically much less than this for such a mission.  Indeed, (Reference \citenum{udrea2012sensitivity}) assume position and velocity standard deviations at the start of the Osiris Rex TAG maneuver of 1cm and 0.1cm/s respectively. The large range of initial conditions demonstrates the ability to learn a policy over a large theater of operations, which has the potential to simplify mission planning. The lander targets a position on the asteroid's pole that is a distance of 250m from the asteroid center. Due to the range of environmental parameters tested, the effect would be identical if the target position was on the equator, or anywhere else 250m from the asteroid's center of rotation. For purposes of computing the Coriolis and centrifugal forces, we translate the lander's position from the target centered reference frame to the asteroid centered reference frame.  We define a landing plane with a surface normal depending on the targeted landing site, which allows use of the Mars landing environment with minimal changes.

At the start of each episode, the lander's mass is uniformly set to between 450 and 500kg. At the start of each episode, the asteroid's angular velocity ($\omega$), mass ($M$), and the local solar radiation pressure (SRP) are uniformly drawn from within the bounds given in Table \ref{tab:ast_forces}. 

\begin{table}[h]
	\fontsize{8}{8}\selectfont
    \caption{Asteroid Lander Initial Conditions}
   \label{tab:exp1_ic}
        \centering 
   \newcolumntype{R}{>{\raggedleft\arraybackslash}p{1.8cm}}
   \begin{tabular}{l | R | R  } 
       Parameter & min & max \\
       \hline
      Distance (m)      & 900 & 1100\\
      Position Polar Angle  $\theta$ (deg)  &    0 &  45 \\
      Position Azimuth Angle $\phi$ (deg)   &  -180 & 180 \\
      Velocity Heading Error (deg) & -45 & 45 \\
      Magnitude Velocity (cm/s) & 5 & 10 \\
      Lander wet mass (kg) & 450  & 500 \\
   \end{tabular}
\end{table}

\begin{table}[h]
	\fontsize{10}{10}\selectfont
    \caption{Asteroid  Environmental Parameters}
   \label{tab:ast_forces}
        \centering 
   \newcolumntype{R}{>{\raggedleft\arraybackslash}p{1.5cm}}
   \begin{tabular}{l | R | R } 
       Parameter & min & max \\
       \hline
       Rotational Velocity $\omega_{x}$ (rad/s) & $-1\times10^{-3}$ & $1\times10^{-3}$ \\
       Rotational Velocity $\omega_{y}$ (rad/s) & $-1\times10^{-3}$ & $1\times10^{-3}$ \\
       Rotational Velocity $\omega_{z}$ (rad/s) & $-1\times10^{-3}$ & $1\times10^{-3}$ \\
       SRP $x$ ($m/s^{2}$) & $-1\times10^{-6}$ & $1\times10^{-6}$   \\
       SRP $y$ ($m/s^{2}$) & $-1\times10^{-6}$ & $1\times10^{-6}$   \\
       SRP $z$ ($m/s^{2}$) & $-1\times10^{-6}$ & $1\times10^{-6}$   \\
       Mass ($\times10^{10}$ kg) & 2 & 20
   \end{tabular}
\end{table}

\subsection{Dynamics Model}

We model the Asteroid landings in 3-DOF, where the translational motion is modeled as follows:

\begin{subequations}
\begin{align}
	\Dot{\mathbf r} &= {{\mathbf v}}\label{eq:EQOM2a}\\
	\Dot{\bf v} &= \frac{{\bf T}}{m} + \mathbf{a}_\text{SRP} - \frac{MG\mathbf{r}_{a}}{\|\mathbf{r}_{a}\|^3}+2\mathbf{\dot{r}}_{a}\times\bm{\omega}_a + (\bm{\omega}_a\times\mathbf{r}_{a})\times\bm{\omega}_a\label{eq:EQOM2b}\\
	\Dot{m} &= -\frac{\lVert{{\bf T}}\rVert}{I_\text{sp}g_\text{ref}} \label{eq:EQOM2c}
\end{align}
\end{subequations}
Here $\mathbf{r}$ is the lander's position in the target centered reference frame, $\mathbf{r}_a$ is the lander's position in the asteroid centered reference frame, ${{\bf T}}$ is the lander's thrust vector,  $g_\text{ref}=9.8$ $\text{m}/\text{s}^{2}$,  $I_\text{sp}=225$ s, and the lander's mass is $m$.  $M$ is the asteroid's mass, and $G$ is the gravitational constant. ${\bf a}_\text{SRP}$ is a vector representing acceleration due to solar radiation pressure (SRP). $\bm{\omega}$ is a vector of rotational velocities in the asteroid centered reference frame.  The lander's nominal wet mass is 500kg. We assume pulsed thrusters with a thrust capability of 2N along each axis in the target centered reference frame. The equations of motion are integrated using 4\textsuperscript{th} order Runge-Kutta integration with a 2s step size, and the navigation period is 6s. 

\subsection{Reward Function}

The reward function used for the 3-DOF asteroid landing is slightly modified from the Mars landing reward function. We do not use a glideslope reward component in the terminal reward, and the shaping rewards are not piecewise.  Specifically, we use the velocity field equations given in \eqref{eq:ast_vtarg1a} through \eqref{eq:ast_vtarg1b}.  The hyperparameters given in Table \ref{tab:HPS} are also modified, with $\tau_{1}=250\text{ s}$ , $\alpha=\text{1.0}$, and $v_{o}=\text{0.5 m/s}$. The landing bonus limits are also modified, with $r_{lim}=\text{1 m}$ and $v_{lim}=\text{0.2 m/s}$. 

\begin{subequations}
\begin{align}
    {\bf v}_{targ}&=-v_{o}\left(\frac{{\bf \hat{r}}}{\lVert{\bf \hat{r}}\rVert}\right)\left(1-\exp\left(-\frac{t_{go}}{\tau_1}\right)\right)\label{eq:ast_vtarg1a}\\
    t_{go}&=\frac{\lVert{\bf \hat{r}}\rVert}{\lVert{\bf \hat{v}}\rVert}\label{eq:ast_vtarg1b} 
\end{align}
\end{subequations}

The observation used in the 3-DOF asteroid environment is given in Equation \eqref{eq:aobs}, where ${\bf v}_\text{error}={\bf v}-{\bf v}_\text{targ}$:

\begin{equation}
\label{eq:aobs}
{\text{obs}}=\begin{bmatrix}{\bf v}_\text{error}  & t_\text{go} \end{bmatrix}
\end{equation}

\section{Experiments}

In each experiment, we optimize the policy using PPO, and then test the trained policy for 10,000 episodes. We compare the test performance of a DR/DV policy, RL policy, and meta-RL policy where the  recurrent layer is unrolled through 1, 20, and 60 timesteps during the forward pass through the policy and value function networks during training. As a shorthand, we will refer to a policy where the recurrent network is unrolled T  steps during the forward pass as a T-step meta-RL policy, and a non-recurrent policy as an RL policy.  Note that unrolling the recurrent network layer for more timesteps results in a policy that can capture longer temporal dependencies. The DR/DV policy instantiates a DR/DV controller, which is thrust limited in the same way as the RL-agents.

For the experiments using the asteroid environment, we tuned the nominal gravity parameter used in the time to go calculation in the DR/DV guidance law for best performance; it turns out that the optimal setting for this parameter is quite a bit higher than the actual environmental gravity. Table \ref{tab:drdv_base} gives baseline performance for the DR/DV policy under ideal conditions in the asteroid landing environment for comparison in the experiments that follow, giving statistics for the terminal position and velocity. Note that even under ideal conditions, the DR/DV policy results in a terminal velocity that is on the high side.

\begin{table}[h]
	\fontsize{10}{10}\selectfont
    \caption{DR/DV Baseline Performance, Asteroid Environment}
   \label{tab:drdv_base}
        \centering 
   \newcolumntype{R}{>{\raggedleft\arraybackslash}p{1.5cm}}
   \begin{tabular}{l | R | R | R  } 
       \hline
      Value & $\mu$ & $\sigma$ & max \\
       \hline
      Position (m)  &  0.1 & 0.0 & 0.6   \\
      Velocity (cm/s) & 10.8 & 7.1 & 37.8
   \end{tabular}
\end{table}

\subsection{Experiment 1: Mars Landing with Engine Failure}

To test the ability of the recurrent policy to deal with actuator failure, we increase the Mars lander's maximum thrust to 24000N. In a 6-DOF environment, each engine would be replaced by two engines with half the thrust, with reduced thrust occurring when one engine in a pair fails. At the start of each episode, we simulate an engine failure in 3-DOF by randomly choosing to limit the available downrange or crossrange thrust by a factor of 2, and limit the vertical (elevation) thrust by a factor of 1.5. Some episodes occur with no failure; we use a failure probability of 0.5. A real engine would hopefully be more reliable, but we want to optimize with each failure mode occurring often. The goal is to demonstrate a level of adaptability that would not be possible without an adaptive guidance system. 
 
Testing the optimized policies resulted in catastrophic failure for the DR/DV policy, but the performance was similar for the RL and meta-RL policies. Therefore, in order to differentiate the performance between these policies,  we further decreased the available thrust during  engine failure, limiting the available downrange or crossrange thrust by a factor of 2.5, with the vertical (elevation) thrust remaining reduced by a factor of 1.5. We see in Tables \ref{tab:exp1_pos} and Tables \ref{tab:exp1_vel} that the DR/DV policy fails catastrophically, whereas the RL policy comes close to a safe landing except for rare outliers that bring the touchdown velocity unacceptably high.  The 1-step Meta-RL policy's performance is acceptable on average, but the outliers are worse than that of the RL policy. Both the 20-step and 60-step Meta-RL policies achieve a safe landing, with minimum terminal glideslope of 87 degrees and fuel consumption similar to that obtained without engine failure.

\begin{table}[h]
	\fontsize{10}{10}\selectfont
    \caption{Experiment 1: Norm of Terminal Position (m)}
   \label{tab:exp1_pos}
        \centering 
   \newcolumntype{R}{>{\raggedleft\arraybackslash}p{0.8cm}}
   \begin{tabular}{l | R | R | R} 
       Statistic & $\mu$ & $\sigma$ & max  \\
       \hline
      DR/DV   &  144.9 & 227.5 & 1322.9   \\
      RL       &  0.7 & 0.3 & 2.2  \\
      Meta-RL 1 step & 0.9 & 6.4 & 511.2   \\
      Meta-RL 20 step &  0.3 & 0.2 & 1.2   \\
      Meta-RL 60 steps & 0.3 & 0.2 & 1.3     \\
   \end{tabular}
\end{table}

\begin{table}[h]
	\fontsize{10}{10}\selectfont
    \caption{Experiment 1: Norm of Terminal Velocity (m/s)}
   \label{tab:exp1_vel}
        \centering 
   \newcolumntype{R}{>{\raggedleft\arraybackslash}p{0.8cm}}
   \begin{tabular}{l | R | R | R } 
       Statistic & $\mu$ & $\sigma$ & max  \\
       \hline
      DR/DV     & 19.51  & 19.90  & 63.13  \\
      RL       &  1.00 & 0.62 &4.76  \\
      Meta-RL 1 step &  0.96 & 0.74 & 42.88 \\
      Meta-RL 20 step & 0.98 & 0.06 & 1.15   \\
      Meta-RL 60 steps & 1.00 & 0.06 & 1.16  \\
   \end{tabular}
\end{table}

\subsection{Experiment 2: Mars Landing with High Mass Variation}

Here we divide the lander engine's specific impulse by a factor of 6, which increases fuel consumption to around 1200kg on average, with a peak of 1600kg. This complicates the guidance problem in that the mass varies by a significant fraction of the lander's initial mass during the descent, and we do not give the agent access to the actual mass during the descent. Although we are using a Mars landing environment for this task, the large variability in mass would be more similar to the problem encountered in an EKV interception of an ICBM warhead, where there is a high ratio of wet mass to dry mass.  

We see in Tables \ref{tab:exp2_pos} and \ref{tab:exp2_vel} that the DR/DV policy has a rather large maximum position error, and an unsafe terminal velocity.  The RL policy and 1-step Meta-RL policy give improved performance, but still result in an unsafe landing. The 20-step meta-RL policy achieves a good landing, which is slightly improved on by the 60-step meta-RL policy.

\begin{table}[h]
	\fontsize{10}{10}\selectfont
    \caption{Experiment 2: Norm of Terminal Position (m)}
   \label{tab:exp2_pos}
        \centering 
   \newcolumntype{R}{>{\raggedleft\arraybackslash}p{0.8cm}}
   \begin{tabular}{l | R | R | R} 
       Statistic & $\mu$ & $\sigma$ & max  \\
       \hline
      DR/DV   & 0.4 & 1.5 & 19.6    \\
      RL      & 0.7 & 0.2 & 3.7 \\
      Meta-RL 1 step & 0.4 & 0.1 & 0.8     \\
      Meta-RL RNN 20 steps &  0.6 & 0.1 & 1.0   \\
      Meta-RL 60 steps &  0.6 & 0.2 & 1.1   \\
   \end{tabular}
\end{table}

\begin{table}[h]
	\fontsize{10}{10}\selectfont
    \caption{Experiment 2: Norm of Terminal Velocity (m/s)}
   \label{tab:exp2_vel}
        \centering 
   \newcolumntype{R}{>{\raggedleft\arraybackslash}p{0.8cm}}
   \begin{tabular}{l | R | R | R } 
       Statistic & $\mu$ & $\sigma$ & max  \\
       \hline
      DR/DV     &  0.63 & 0.96 & 5.39    \\
      RL        & 0.92 & 0.26 & 5.25     \\
      Meta-RL 1 steps & 0.98 & 0.42 & 6.48    \\
      Meta-RL 20 step & 1.17  & 0.07 & 1.36   \\
      Meta-RL 60 steps &  1.06 & 0.05 & 1.21  \\
   \end{tabular}
\end{table}

\begin{figure}[h]
\begin{center}
\includegraphics[width=.9\linewidth]{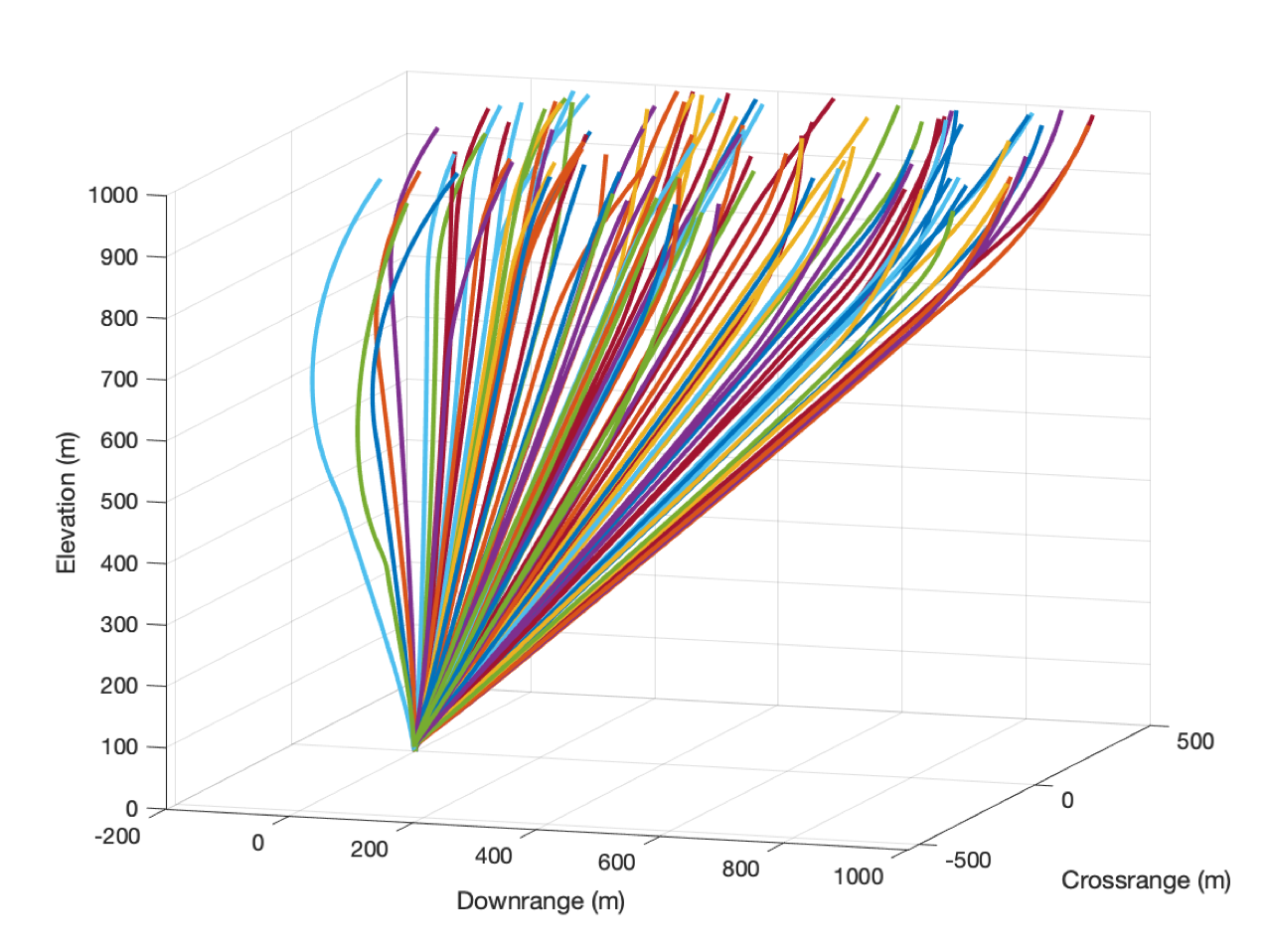}
\caption{Experiment 3: Landing with Integrated Guidance and Navigation. Trajectories are samples from neural network policy in closed-loop.}
\label{fig:exp3_trajectory}
\end{center}
\end{figure}

\subsection{Experiment 3: Mars Landing using Radar Altimeter Observations}

In this task, the observations are simulated Doppler altimeter readings from the lander using a digital terrain map (DTM) of the Mars surface in the vicinity of Uzbois Valis. Since the simulated beams can cover a wide area of terrain, we doubled the map size by reflecting the map and joining it with original map. Note that the agent does not have access to the DTM, but will learn how to successfully complete a maneuver using rewards received from the environment. Although these observations are a function of both lander position and lander velocity, the observations do not satisfy the Markov property as there are multiple ground truth positions and velocities that could correspond to a given observation, making the optimal action a function of the history of past altimeter readings. For the simulated altimeter readings, the agent's state in the target centered reference frame is transformed to the DTM frame, which has a target location of [4000, 4000, 400] meters.

In order to simulate altimeter readings fast enough to allow optimization to complete in a reasonable time, we had to create a fast altimeter model. The model uses a stack of planes with surface normals in the $z$ (elevation direction) that spans the elevations in the DTM.  Each of the four radar beams has an associated direction vector, which, in conjunction with the lander's position, can quickly be used to find the intersection of the beam and the planes.  The intersection $x$,$y$ indices are used to index the DTM, and the plane intersection with a z value closest to that of the indexed DTM elevation is used to determine the distance between the lander and the terrain feature hit by the beam.  This is extremely fast (about 1000X faster than the ray-casting approach we used in  [\citenum{gaudet2014navigation}], but is error prone at lower elevations as sometimes the closest distance between DTM elevation and associated plane intersect $z$ component is the far side of a terrain feature. Rather than call this a bug, we use it to showcase the ability of a recurrent policy to get remarkably close to a good landing, given the large errors.  The reduction in accuracy at lower elevations is apparent in Table \ref{tab:mm_acc}. The accuracy was estimated by choosing 10,000 random DTM locations and casting a ray to a random position at the designated elevation. The length of this ray is the ground truth altimeter reading.  We then checked what the measurement model returned from that lander position, with the error being the difference.  Note that the DTM elevations range from 0 to 380m.  In this scenario, the lander target's a terminal position 50m above the top of a hill at 350m elevation.

\begin{figure}[h]
\begin{center}
\includegraphics[width=.9\linewidth]{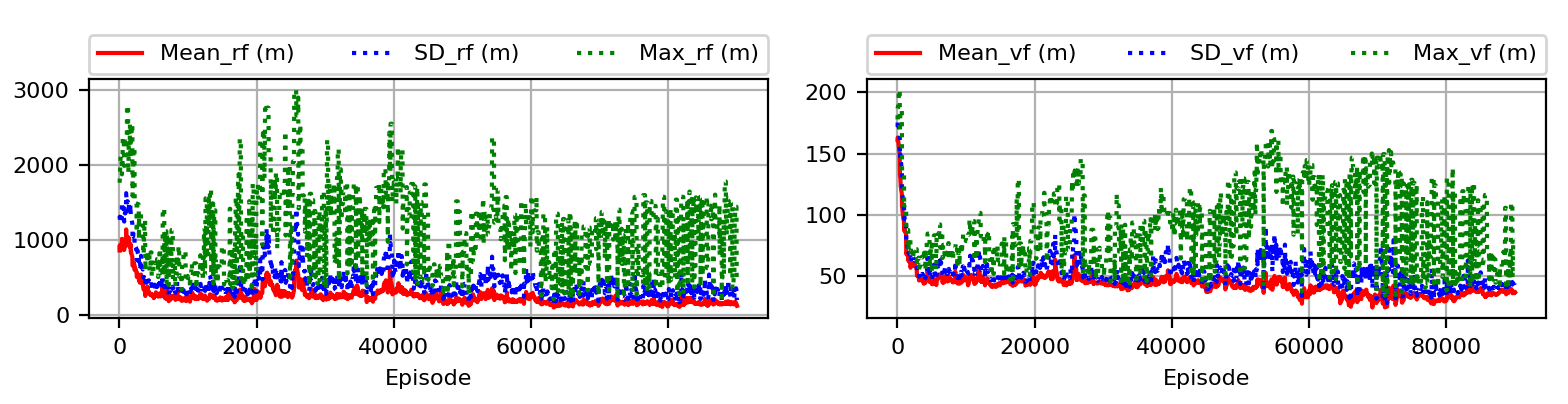}
\caption{Experiment 3: Learning Curves for 1 step RNN}
\label{fig:exp2_lc-1step}
\end{center}
\end{figure}

\begin{figure}[h]
\begin{center}
\includegraphics[width=.9\linewidth]{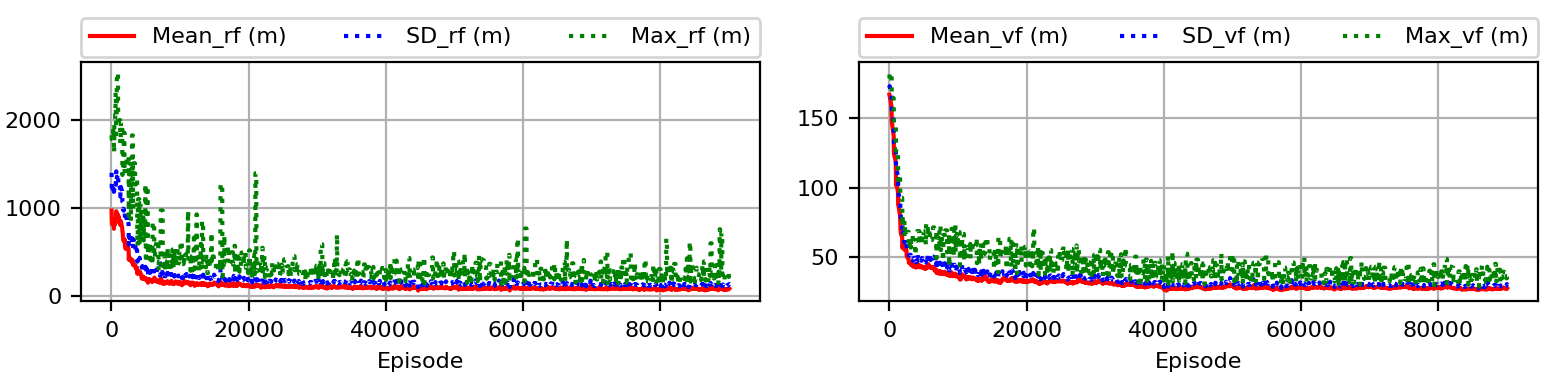}
\caption{Experiment 3: Learning Curves for 120 step RNN}
\label{fig:exp2_lc}
\end{center}
\end{figure}

The altimeter beams are modeled as having equal offset angles ($\pi/8$ radians) from a direction vector that points along the lander's body frame negative $z$-axis.  The lander's negative z-axis is then aligned at each simulation step in a direction that is averaged between the lander's velocity vector and straight down. We thought this a reasonable assumption as we are modeling the lander in 3-DOF. We see from Tables \ref{tab:exp3_pos} and \ref{tab:exp3_vel} that although you would not want to entrust an expensive lander to this integrated guidance and navigation algorithm, the performance is remarkably good given the altimeter inaccuracy at lower elevations. We also note a steady improvement in performance as the number of recurrent steps is increased. Learning curves for the 1-step and 20-step RNN's are shown in Figures \ref{fig:exp2_lc-1step} and \ref{fig:exp2_lc}, which plots statistics for terminal  position ($r_f$) and terminal velocity ($v_f$) as a function of episode, with the statistics calculated over the 30 episodes used to generate rollouts for updating the policy and value function.  We see from the learning curves that the amount of steps we unroll the recurrent network  in the forward pass has a large impact on optimization performance, and that for the 120 step case, the optimization initially makes good progress, but then stalls, likely due to the highly inaccurate altimeter readings at lower altitudes. Figure \ref{fig:exp3_trajectory} shows sampled trajectories over a 4 km squared deployment area.

\begin{table}[h]
	\fontsize{8}{8}\selectfont
    \caption{Experiment3: Altimeter Error as function of lander elevation (m)}
   \label{tab:mm_acc}
        \centering 
   \newcolumntype{R}{>{\raggedleft\arraybackslash}p{1.8cm}}
   \begin{tabular}{l | l | l | l | l } 
       Elevation (DTM Frame) & mean (m) & std (m) & max (m) & miss \% \\
       \hline
       500 & 122 & 528 & 4467 & 12 \\
       600 & 25 & 201 & 2545 & 6 \\
       700 & 8 & 92 & 1702 & 4 \\
       800 & 4 & 60 & 1300 & 2 \\
     
      \hline
    
   \end{tabular}
\end{table}

\begin{table}[h]
	\fontsize{10}{10}\selectfont
    \caption{Experiment 3: Norm of Terminal Position (m)}
   \label{tab:exp3_pos}
        \centering 
   \newcolumntype{R}{>{\raggedleft\arraybackslash}p{0.8cm}}
   \begin{tabular}{l | R | R | R} 
       Statistic & $\mu$ & $\sigma$ & max  \\
       \hline
       RL       &  131 & 101 & 796  \\
      Meta-RL 1 step & 114 & 95 & 1359  \\
      Meta-RL 20 step &  78 & 41 & 390 \\
      Meta-RL 120 steps & 72 & 40 & 349  \\
      Meta-RL 200 steps & 59 & 42 & 288 
   \end{tabular}
\end{table}

\begin{table}[h]
	\fontsize{10}{10}\selectfont
    \caption{Experiment 3: Norm of Terminal Velocity (m/s)}
   \label{tab:exp3_vel}
        \centering 
   \newcolumntype{R}{>{\raggedleft\arraybackslash}p{0.8cm}}
   \begin{tabular}{l | R | R | R } 
       Statistic & $\mu$ & $\sigma$ & max  \\
       \hline
      RL       &  48  & 18 &  145  \\
      Meta-RL 1 step &  37 & 4  & 67  \\
      Meta-RL 20 step &  26 & 2 & 39   \\
      Meta-RL 120 steps &  28 & 2  & 44    \\
      Meta-RL 200 steps & 23 & 4 & 40 \\
   \end{tabular}
\end{table}

In a variation on this experiment, we assume that the lander has the ability to point its radar altimeters such that the central direction vector remains fixed on the target location, and therefore the beams themselves bracket the target.  This functionality could be achieved with phased array radar, but a separate pointing policy would need to be learned that keeps the beams pointed in the required direction.   Here we see (Tables \ref{tab:exp3p_pos} and \ref{tab:exp3p_vel}) that performance markedly improves.  We postulate one reason for the improved performance is that the altimeter beams remain in areas of high terrain diversity.  Indeed, when we repeat the experiment for a landing site further to the south (bottom of DTM), we find that performance degrades. Another factor could be that since the policy is focused on a small portion of the map, it does not "forget" the relationship between observations, actions, and advantages.

\begin{table}[h]
	\fontsize{10}{10}\selectfont
    \caption{Experiment 3 (Target Pointing): Norm of Terminal Position (m)}
   \label{tab:exp3p_pos}
        \centering 
   \newcolumntype{R}{>{\raggedleft\arraybackslash}p{0.8cm}}
   \begin{tabular}{l | R | R | R} 
       Statistic & $\mu$ & $\sigma$ & max  \\
       \hline
       RL       & 3.3  & 1.9 & 108.6  \\
      Meta-RL 1 step & 3.3  & 1.9 & 108.6  \\
      Meta-RL 20 step & 0.3 & 1.2 &  116.0   \\
      Meta-RL 60steps & 0.4  & 1.5 & 111.5   \\
      Meta-RL 120 steps &  0.6 & 1.6 & 139.8 \\
   \end{tabular}
\end{table}

\begin{table}[h!]
	\fontsize{10}{10}\selectfont
    \caption{Experiment 3 (Target Pointing): Norm of Terminal Velocity (m/s)}
   \label{tab:exp3p_vel}
        \centering 
   \newcolumntype{R}{>{\raggedleft\arraybackslash}p{0.8cm}}
   \begin{tabular}{l | R | R | R } 
       Statistic & $\mu$ & $\sigma$ & max  \\
       \hline
      RL       &  4.92 & 2.29 & 84.16  \\
      Meta-RL 1 step &  5.75  & 1.38 & 69.80  \\
      Meta-RL 20 step & 1.61 & 0.60 & 64.84    \\
      Meta-RL 60 steps & 2.00 &  0.92 & 53.73     \\
      Meta-RL 120steps &  2.23 & 0.94 & 57.47 \\
   \end{tabular}
\end{table}

Taking into account the number of large outliers, it is probably best to focus on the average performance when comparing the network architectures.  The general trend is that performance increases as we increase the number of steps we unroll the recurrent layer for the forward pass. This implies that the temporal dependencies for this task probably span a significant fraction of a single episode.

\subsection{Experiment 4: Mars Landing with State Estimation Bias}

In this experiment, the agent has access to a corrupted version of the ground truth state $\mathbf{x}=[\bf r, \bf v]$, specifically the agent's position observation is   $\mathbf{o}_r=\mathbf{r}+p_{r}|\mathbf{r}|$ and the agent's velocity observation is $\mathbf{o}_v=\mathbf{v}+p_{v}|\mathbf{v}|$,  where $p_r$ and $p_v$ are uniformly drawn between -0.1 and 0.1 at the start of each episode.  We expect that the meta-RL policy will be able to quickly adapt to the level of sensor bias in a given episode.

The results are given in Tables \ref{tab:exp4_pos} and \ref{tab:exp4_vel}. Here we do not see any large performance difference between RL and 60-step meta-RL policies. Interestingly, the 1-step and 20-step meta-RL have poor performance. This could be due to the policy misinterpreting the sensor bias as a force acting on the lander. To illustrate, assume a sensor bias of +10\%. For the case of no sensor bias, a given thrust level might result in some change in velocity. But with the sensor bias, this change in velocity would be larger, which could be interpreted as a force acting on the spacecraft in addition to the thrust. By attempting to adapt to this fictitious force, performance might be  impaired rather than improved. There may be information in the temporal dependencies that allow distinguishing between external forces and sensor noise that are only captured with more than 20 steps. Since the RL policy uses only the current observation and advantage to update the policy, it cannot attempt to adapt to the sensor noise, but merely learns a policy that is robust to the noise. This could explain why the RL policy performs much better than the 1 or 20 step meta-RL policies.

\begin{table}[h]
	\fontsize{10}{10}\selectfont
    \caption{Experiment 4: Norm of Terminal Position (m)}
   \label{tab:exp4_pos}
        \centering 
   \newcolumntype{R}{>{\raggedleft\arraybackslash}p{0.8cm}}
   \begin{tabular}{l | R | R | R} 
       Statistic & $\mu$ & $\sigma$ & max  \\
       \hline
      DR/DV     & 0.1 & 1.7 &  75.7  \\
      MLP       &  0.9 & 0.3 & 3.3 \\
      Meta-RL 1 steps & 0.6 & 0.7 & 59.6  \\
      Meta-RL 20 step & 0.6 & 1.9 & 183.8 \\
      Meta-RL 60 steps &  0.5 & 0.4 & 5.5  \\
   \end{tabular}
\end{table}

\begin{table}[h!]
	\fontsize{10}{10}\selectfont
    \caption{Experiment 4: Norm of Terminal Velocity (m/s)}
   \label{tab:exp4_vel}
        \centering 
   \newcolumntype{R}{>{\raggedleft\arraybackslash}p{0.8cm}}
   \begin{tabular}{l | R | R | R } 
       Statistic & $\mu$ & $\sigma$ & max  \\
       \hline
      DR/DV     & 0.48  & 0.93 & 20.14 \\
      MLP        & 1.11  & 0.11 &  1.48 \\
      Meta-RL 1 steps & 1.11 &  0.24 & 22.39  \\
      Meta-RL 20 step & 1.40 &   0.25 & 22.26  \\
      Meta-RL 60 steps & 1.31 & 0.12 & 1.69  \\
   \end{tabular}
\end{table}

\subsection{Experiment 5:  Asteroid Landing with Unknown Dynamics}

This experiment is a simulated landing on an asteroid with unknown and highly variable environmental dynamics, where the goal is to land on the asteroid's pole, which is assumed to be 250m from the asteroid center for purposes of computing centrifugal and Coriolis forces. We chose an asteroid landing environment for this task because the asteroid's rotation can cause the  forces acting on the lander to vary widely, creating in effect an environment with unknown dynamics.

Tables \ref{tab:exp5_pos} and \ref{tab:exp5_vel} give the test results. Again, we see that the DR/DV policy gives unacceptable performance, with a large position error and a landing velocity close to 1m/s. Interestingly, there is no real difference between the RL policy and the meta-RL policies. For this scenario, it appears that optimizing using parameter uncertainty provides the same performance as an adaptive policy.  It is worth mentioning that we did perform a variation of this experiment in the Mars landing environment where the acceleration due to gravity was randomly drawn between -4m/s and 4m/s at the start of each episode.  In this highly unrealistic environment, only the adaptive policies where the recurrent layers were unrolled for at least 20 steps during optimization consistently achieved a safe landing.

\begin{figure}[h]
\begin{center}
\includegraphics[width=.9\linewidth]{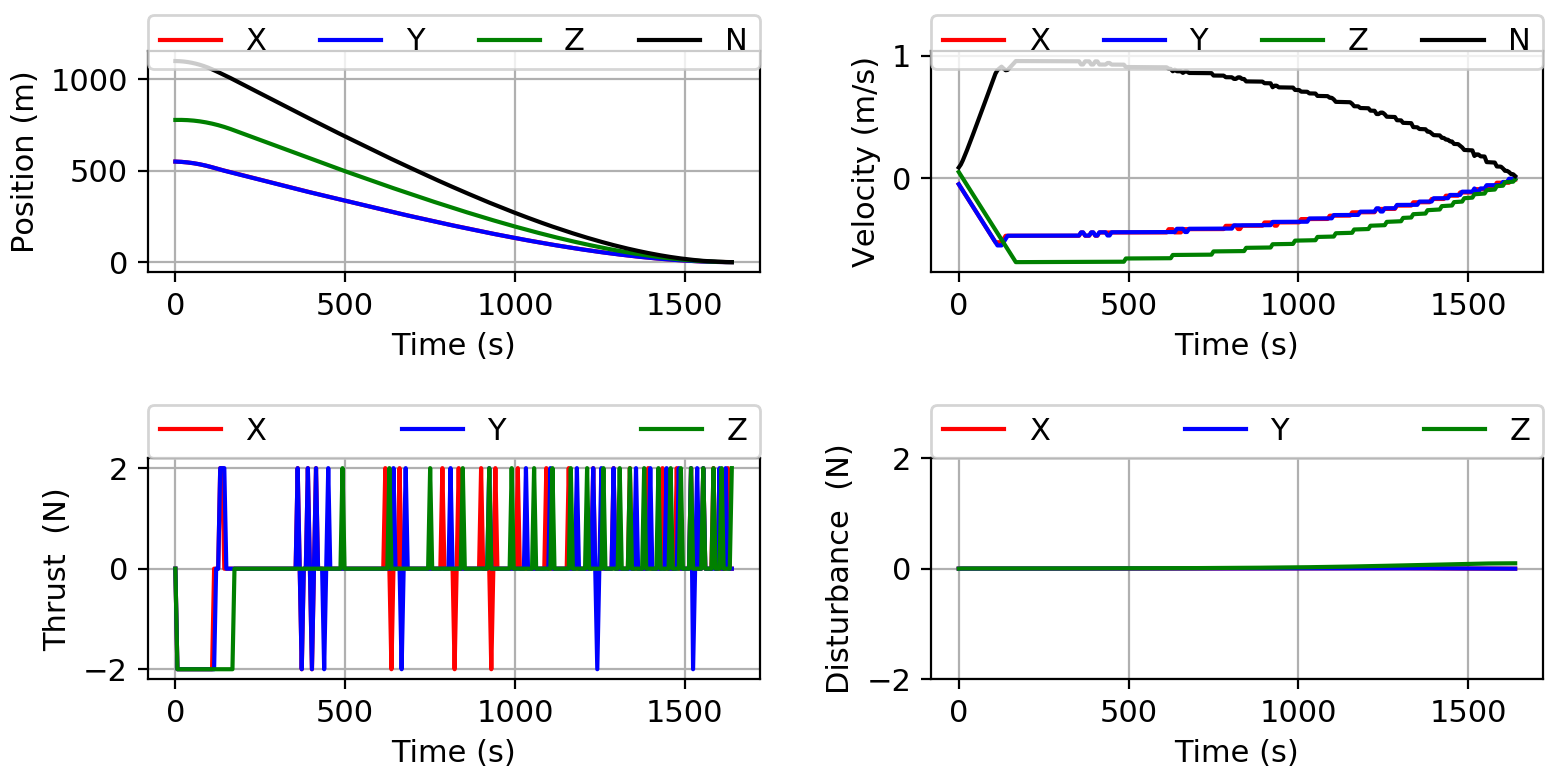}
\caption{Experiment 5: 60-step Meta-RL Policy with low disturbance}
\label{fig:exp5_lowdist}
\end{center}
\end{figure}

\begin{figure}[h!]
\begin{center}
\includegraphics[width=.9\linewidth]{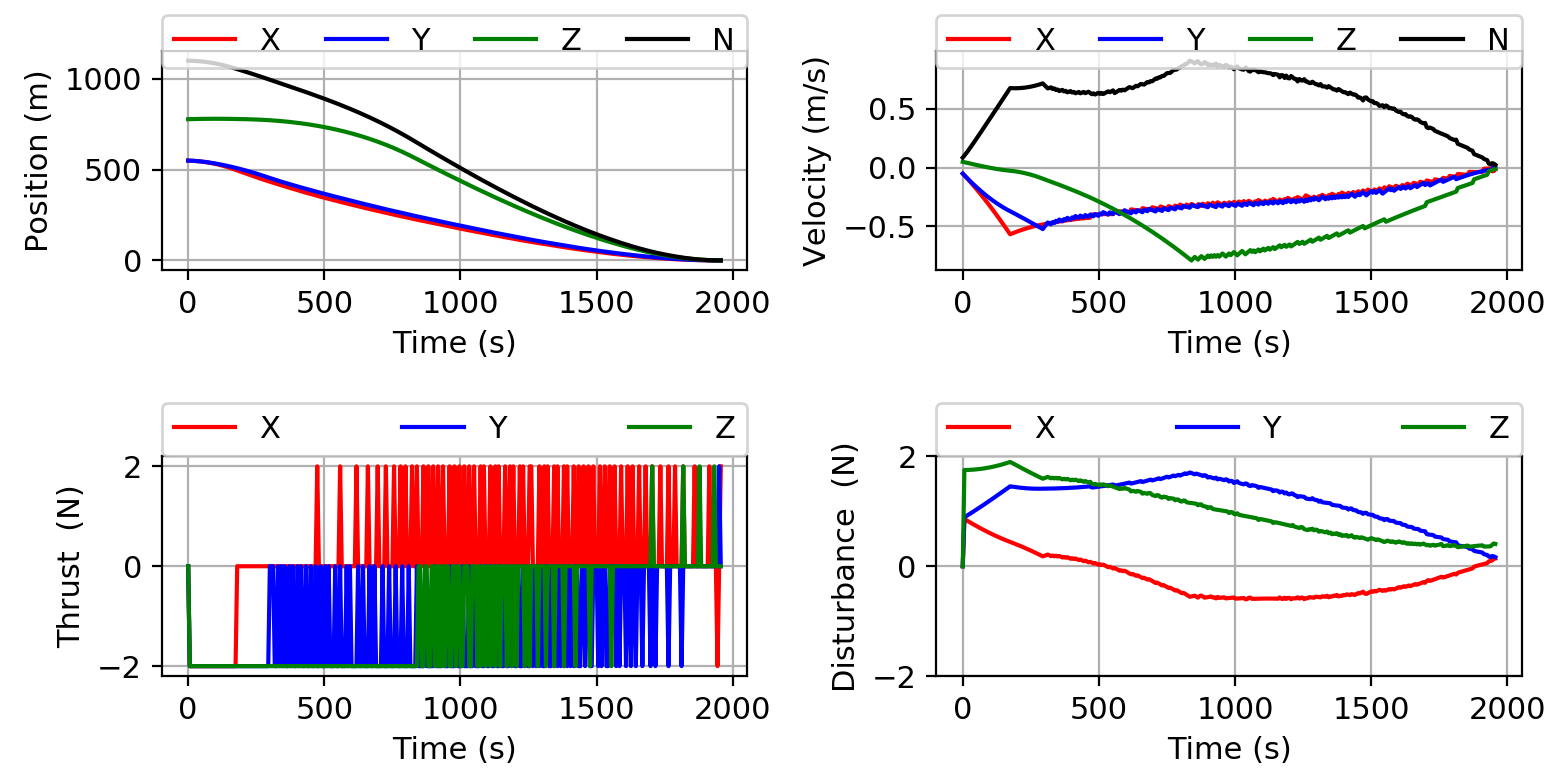}
\caption{Experiment 5: 60-step Meta-RL Policy with high disturbance}
\label{fig:exp5_highdist}
\end{center}
\end{figure}

Figures \ref{fig:exp5_lowdist} and \ref{fig:exp5_highdist} illustrate a single trajectory from the 60-step Meta-RL policy with a low and high external disturbance, respectively. The X, Y, and Z curves are elevation, crossrange, and elevation in the target centered reference frame. The N curve is the L2 norm. Note that the disturbance approaches the lander's maximum thrust capability in the high disturbance case.

\begin{table}[h]
	\fontsize{10}{10}\selectfont
    \caption{Experiment 5: Norm of Terminal Position (m)}
   \label{tab:exp5_pos}
        \centering 
   \newcolumntype{R}{>{\raggedleft\arraybackslash}p{0.8cm}}
   \begin{tabular}{l | R | R | R} 
       Statistic & $\mu$ & $\sigma$ & max  \\
       \hline
      DR/DV     & 0.3 & 1.9  &  44.0  \\
      RL       & 0.1 & 0.0 & 0.3 \\
      Meta-RL  1 steps & 0.1 & 0.1 & 0.5  \\
      Meta-RL  20 step & 0.1 & 0.1 & 0.5 \\
      Meta-RL  60 steps & 0.1 & 0.1 & 0.6  \\
   \end{tabular}
\end{table}

\begin{table}[h]
	\fontsize{10}{10}\selectfont
    \caption{Experiment 5: Norm of Terminal Velocity (cm/s)}
   \label{tab:exp5_vel}
        \centering 
   \newcolumntype{R}{>{\raggedleft\arraybackslash}p{0.8cm}}
   \begin{tabular}{l | R | R | R } 
       Statistic & $\mu$ & $\sigma$ & max  \\
       \hline
      DR/DV     & 7.5  & 7.6  & 85.7  \\
      RL        & 2.1 & 0.7 & 5.1 \\
      Meta-RL 1 steps & 2.6 & 0.8 & 16.9  \\
      Meta-RL 20 step & 2.5 & 0.9 &  7.2  \\
      Meta-RL 60 steps & 2.5 & 0.9 & 7.9  \\
   \end{tabular}
\end{table}

\subsection{Experiment 6:  Asteroid Landing using Doppler LIDAR altimeter readings}

In our final experiment, we integrate navigation and guidance in the Asteroid 3-DOF environment, giving the agent access to only Doppler LIDAR altimeter readings during the landing. We simplify the problem by using a reduced set of initial conditions, with the polar angle in Table \ref{tab:exp1_ic} reduced to 22.5 degrees, and the range of rotational velocities is reduced to between $-1\times10^{-5} \text{ and } 1\times10^{-5}$ rad/s. We assume that in the body frame one altimeter beam is pointed straight down, and four more beams are equally spaced with an offset angle of 12 degrees from the first. Note that this could be implemented with a single scanning LIDAR system. At each simulation time step, we then rotate the body so that the -z body frame axis is aligned with the lander's velocity vector, i.e., the center beam is always aligned with the lander's velocity vector.   

A shape model of asteroid rq-36 with 1-m resolution is down sampled to 5m resolution to reduce the required amount of ray-casting operations, and we use the same ray casting approach as in [\citenum{gaudet2014real}] to determine where the altimeter beams intersect the asteroid and the Doppler closing velocity associated with each beam. The observation given to the agent is then the concatenation of the five range readings and five Doppler closing velocities $\mathbf{o} \in \mathbb{R}^{10}$. 

Tables \ref{tab:exp6_pos} and \ref{tab:exp6_vel} tabulate the test results.  We see that the average miss distance approaches the shape model resolution, although the occasional outlier exceeds 3 times the resolution. Terminal velocity is also on the high side. We see a similar patter to that in the Mars landing using RADAR experiment, where the best results are for the 60-step meta-RL policy, and a steady decrease in performance as we unroll the recurrent layer fewer steps in the forward pass during optimization.

\begin{table}[h]
	\fontsize{10}{10}\selectfont
    \caption{Experiment 6: Norm of Terminal Position (m)}
    \label{tab:exp6_pos}
        \centering 
   \newcolumntype{R}{>{\raggedleft\arraybackslash}p{0.8cm}}
   \begin{tabular}{l | R | R | R} 
       Statistic & $\mu$ & $\sigma$ & max  \\
       \hline
      RL       &  8.4 & 22.5 & 1533.2\\
      Meta-RL  1 steps & 44.6 & 167.4 & 1018.0  \\
      Meta-RL  20 step & 9.8 & 5.9 & 36.8\\
      Meta-RL  60 steps & 7.9 & 4.6 & 28.2 \\
   \end{tabular}
\end{table}

\begin{table}[h]
	\fontsize{10}{10}\selectfont
    \caption{Experiment 6: Norm of Terminal Velocity (cm/s)}
   \label{tab:exp6_vel}
        \centering 
   \newcolumntype{R}{>{\raggedleft\arraybackslash}p{0.8cm}}
   \begin{tabular}{l | R | R | R } 
       Statistic & $\mu$ & $\sigma$ & max  \\
       \hline
      RL        & 13.7 & 5.1 & 312.3\\
      Meta-RL 1 steps &  30.7 & 77.4 & 42.2\\
      Meta-RL 20 step & 22.8 & 8.2 & 54.9  \\
      Meta-RL 60 steps & 14.6 & 3.0 & 35.7  \\
   \end{tabular}
\end{table}

\section{Conclusion}

We optimized an adaptive guidance system using reinforcement meta learning with recurrent policy and value function in four different environments, and demonstrated the ability of the optimized policy to adapt in real time. In all cases both the RL and 60-step Meta-RL policies outperformed the DR/DV policy, and in the engine failure and high mass variation experiments, the 60-step meta-RL policy outperformed the RL policy.  In some cases the 1-step and  20-step Meta-RL policies performed relatively poorly.  For a given task, there is likely an optimal number of steps to unroll the recurrent layer during the forward pass, with the optimal number of steps depending on the bandwidth of the temporal dependencies in the sequences of observations, actions, and advantages.  If this number of steps is too low  the policy may not learn dependencies that are critical to the mastering the task. Although not examined in this work, too many steps could also be a problem if it is much greater than the temporal dependency bandwidth, and when the number of steps gets very large, there are also issues with exploding and vanishing gradients. Therefore, we suggest that the number of steps to unroll the recurrent network in the forward pass should be treated as a hyperparameter during optimization. 

We then optimized an integrated guidance and navigation system for both the 3-DOF Mars and asteroid landing environments, with the agent only getting access to Doppler radar and LIDAR altimeter readings, respectively. In both of these cases, the recurrent layer of the Meta-RL policy provided superior performance to the RL policy.  This was expected, as both of these tasks use non-Markov observations, with a given ground truth state potentially mapping to multiple observations. Only by looking at a history of observations is it possible to reduce this to a one to one mapping. Since sensor output feedback is an open problem with optimal control methods, we did not compare performance to that of the DR/DV guidance law. The results of the Mars landing experiment were rather poor, which could be at least partly attributed to the errors in the fast ray tracing algorithm that was used to model the sensor.  The asteroid landing experiment produced better results, closer to that which would be obtainable with access to the ground truth lander state. We attribute the higher performance in the asteroid landing environment to two factors: an accurate ray tracing algorithm and the curvature of the asteroid, the latter resulting in a given ground truth state mapping to fewer observations. 

The take away from this work is that the ability to optimize using parameter uncertainty leads  to robust policies, and the ability of a recurrent policy to adapt in real time to dynamic environments makes it the best performing option in environments with unknown or highly variable dynamics. This capability has the potential to dramatically simplify mission planning for asteroid close proximity missions. The results for integrating guidance and navigation were not as promising, as performance was poor compared to what is possible using a separate navigation system. To illustrate, in [\citenum{gaudet2014real}], we couple a navigation system using a Rao-Blackwellized particle filter and LIDAR measurements with the DR/DV policy (in an environment with known dynamics) and consistently achieve a simulated pinpoint soft landing. We repeat this for a Mars landing using RADAR altimeter measurements in  [\citenum{gaudet2014navigation}]. Clearly, more work is needed to develop a fully integrated guidance, navigation, and control system using the RL framework that achieves higher performance than that possible with a separate navigation system.

We only scratched the surface of applications that would benefit from adaptive guidance, and future work will use the RL meta-learning framework for problems in orbital refueling, exoatmospheric intercepts, and hypersonic reentry. We will also explore different sensor models when integrating navigation into the policy, where observations will consist of direct sensor output using simulated camera images, flash LIDAR, and electo-optical sensors.

\section{References}

\bibliographystyle{AAS_publication}   
\bibliography{Acta2019}   

\end{document}